\documentclass[12pt,twoside,fleqn]{article}
\usepackage{epsfig}
\usepackage{psfig}
\makeatletter
\@addtoreset{equation}{section}
\makeatother

\begin{document}
\thispagestyle{empty}
%
 \mbox{} \hfill FSU-SCRI-97-113\\
 \mbox{} \hfill BI-TP 97/36\\
 \mbox{} \hfill September 1997\\
\begin{center}
\vspace*{1.0cm}
{{\Large \bf      The Gluon Propagator at High Temperature: \\
                  Screening, Improvement and \\
\vspace{0.8ex}    Non-Zero Momenta }} \\
\vspace*{1.0cm}
{\large U. M. Heller$^1$, F. Karsch$^2$ and J. Rank$^{1,2}$}
\\
\vspace*{1.0cm}
{\normalsize
$\mbox{}$ {
$^1$ SCRI, Florida State University, Tallahassee, FL 32306-4130, USA \\
$^2$ Fakult\"at f\"ur Physik, Universit\"at Bielefeld,
P.O. Box 100131, D-33501 Bielefeld, Germany
}}\\
\vspace*{2.0cm}
{\Large \bf Abstract}
\end{center}
\setlength{\baselineskip}{1.3\baselineskip}

We study the gluon propagator and the singlet potential in Landau gauge in
the deconfined phase of SU(2) lattice gauge theory, using both the standard
Wilson action and a tree-level Symanzik improved action. From the long-distance
behavior of correlation functions of temporal and spatial components of
the gauge fields we extract electric ($m_e$) and magnetic ($m_m$)
screening masses. For the magnetic mass we find
$m_m(T) = 0.456(6) \, g^2(T) \, T$. The electric mass can be described by a
next-to leading order ansatz, obtained from one loop resummed perturbation
theory. However, the best description is given by
$m_e(T) = \sqrt{1.70(2)} \, g(T) \, T$. The electric screening mass thus is
different from its lowest order perturbative prediction even for
temperatures as high as $T \sim 10^4 \, T_c$.

\newpage
\setcounter{page}{1}

\section{Introduction}
One of the main features of the high temperature plasma phase of QCD is the
occurrence of chromo-electric and -magnetic screening masses ($m_e$ and $m_m$)
which control the infrared behavior of the theory \cite{Lin80}.  
The electric screening mass, which has been known in leading order perturbation
theory for some time ($m_e \sim gT$), gives rise to the Debye screening
of the heavy quark potential. The magnitude of $m_e$ influences strongly 
the existence or non-existence of hadronic bound states in the high temperature
phase. An understanding of its temperature dependence is therefore essential
for any further analysis of the quasi-particle excitation spectrum in the
QCD plasma phase. Also, a non-vanishing magnetic mass, which is entirely
of non-perturbative origin and generally is expected to
be ${\cal O} (g^2 T)$, does contribute in next-to-leading order 
to $m_e$ \cite{Re93,Re94}. Beyond leading order perturbation theory one thus
needs a non-perturbative analysis also for the quantitative determination
of the electric mass\footnote{The non-perturbative structure of the electric
mass beyond leading order perturbation theory has recently also been 
discussed in \cite{BlIa96}. The interesting possibility has been raised there
that non-perturbative effects may arise even without generation of a
non-vanishing magnetic mass.}.

The definition of gluonic screening masses is not without problems.  
In general the masses are related to the low momentum behavior of the gluon 
propagator, $\Pi_{\mu \mu} (p_0, \vec{p})$. However, as the gluon propagator 
itself is a gauge dependent quantity it is not obvious that a ``gluon mass'' 
extracted from it would have a physical interpretation. Using the zero
momentum limit ($|\vec{p}|\rightarrow 0$) in the static sector
($p_0 \equiv 0$) of $\Pi_{\mu \mu}$ does not yield a gauge invariant definition
of the screening masses. It has, however, been shown \cite{KoKuRe90,KoKuRe91}
that the pole masses defined through
$m_\mu^2 = \Pi_{\mu \mu} (0, |\vec{p}|^2=-m_\mu^2)$ are, within a wide class
of gauges, gauge invariant to arbitrary order in perturbation theory. Exactly
these pole masses are obtained from the exponential decay of finite temperature
gluon correlation functions at large spatial separations. Although these
correlation functions have to be calculated in a fixed gauge the pole masses
extracted from them will be gauge independent.
Alternatively one may insist on defining gluon screening masses through
gauge invariant operators, e.g. Polyakov loop correlation functions, which
are related to the heavy quark potential at finite temperature. Observables,
which project onto states with the correct quantum numbers have been discussed
in Ref.~\cite{ArYa95} and have recently been used to analyse the high
temperature phase of QCD within the framework of dimensional reduction
\cite{KaLaRuSh97,KaLaPeRaRuSh97}. In how far the definition of a screening
mass through gauge invariant operators agrees with the pole mass has to be
further analysed. The experience made with similar concepts in gauge-Higgs
models shows that there is the possibility that gauge invariant operators only
project onto a superposition of several elementary gluon excitations
\cite{KaNePaRa96}.

In this paper we analyse gauge dependent gluon correlation functions in 
Landau gauge. We have studied gluon correlation functions at finite
temperature previously for the SU(2) gauge theory in a limited temperature
interval  ($1 < T/T_c < 20$) \cite{HeKaRa95}. This analysis will be extended
here in various ways. We have calculated magnetic and electric screening masses
at much higher temperatures, up to $T \!\sim\! 10^4 \, T_c$, in order to
analyse the approach to the perturbative regime. Furthermore, we study this
time the dependence of the gluon correlation functions on momenta. In 
addition, we now also analyse the color singlet heavy quark potential, which
at high temperatures is dominated by single gluon exchange. Moreover, we have
also used a tree-level Symanzik improved gauge action in addition to the
standard Wilson action in order to get some control over systematic lattice
discretization errors.

In the next section we briefly discuss the observables we are going to analyse.
In section~\ref{sec:temperature} we define the actions we have used in our
simulations and describe the procedure followed to fix
the relation between bare gauge couplings and the temperature. The
determination of electric and magnetic screening masses from gluon correlation 
functions is discussed in section~\ref{sec:screening_masses}. In 
section~\ref{sec:polyakov} we discuss the determination of the electric 
screening mass from the color singlet heavy quark potential. Furthermore we
describe the influence of the improvement of the action on this potential.
Finally, we give our conclusions in section~\ref{sec:summary}.

\section{Gluon Propagator on the Lattice}
\label{sec:gluonprop}
The main purpose of this investigation is an analysis of the gluon
propagator at small momenta. We will extract screening masses from correlation
functions of gauge fields $A_\mu (x_0,\vec{x})$ at large spatial 
separations. We will analyse dispersion relations for these screening
masses in the static ($p_0\equiv 0$) sector. For this purpose we introduce
momentum dependent gauge fields,
\begin{equation}
\tilde{A}_\mu (p_\bot,x_3)  \,=\! \sum_{x_0, x_\bot} e^{i \, x_\bot p_\bot}
A_\mu(x_0,x_\bot,x_3) \quad ,
\label{atilde}
\end{equation}
and the corresponding correlation functions 
\begin{equation}
\tilde{G}_\mu(p_\bot,x_3)  = \left\langle \mbox{Tr} \; \tilde{A}_\mu (p_\bot,x_3) 
\tilde{A}_\mu^\dagger (p_\bot,0) \right\rangle
\label{gtilde}
\end{equation}
with $x_\bot = (x_1,x_2)$ and $p_\bot = (p_1,p_2)$. On a finite lattice, the
momenta are given by $p_i = 2 \pi k_i / (a N_i)$, with
$k_i = - \frac{1}{2} N_i + 1,\ldots,\frac{1}{2} N_i$ and $N_i$ being the
length of the lattice in the $i$-th direction.

The long-distance behavior of $\tilde{G}$ yields the energy in the electric
and magnetic sector respectively, i.e.\footnote{Note that $\tilde{G}_3(p_\bot,x_3)$ is
independent of $x_3$ in Landau gauge.}
\begin{eqnarray}
G_e(p_\bot,x_3) & \!\equiv\! & \tilde{G}_0(p_\bot,x_3) \nonumber \\
& \,\sim\, & \exp\{-E_e(p_\bot) \, x_3\} \quad \mbox{for} \; x_3 \gg 1 ,
\nonumber \\
G_m(p_\bot,x_3) & \!\equiv\! & \frac{1}{2}
\left( \tilde{G}_1(p_\bot,x_3) + \tilde{G}_2(p_\bot,x_3) \right)
\label{eq:gluonfunctions} \\
& \!\sim\! & \exp\{-E_m(p_\bot) \, x_3\} \quad \mbox{for} \; x_3 \gg 1 .
\nonumber
\end{eqnarray}
For $p_\bot\equiv (0,0)$ the long-distance behavior of these correlation
functions thus defines electric and magnetic screening masses, which are
related to the gluon polarization tensor,
\begin{equation}
m_\mu^2 = \Pi_{\mu \mu} (0, \vec{p}\,^2 = -m_\mu^2) \quad .
\label{scrmass}
\end{equation}
Using this definition, the leading correction to the lowest order perturbation
theory result for the electric screening mass,
\begin{equation}
m_{e,0}(T) = \sqrt{\frac{2}{3}} \: g(T) \, T \quad,
\label{lowestscreen}
\end{equation}
can be calculated in one-loop resummed perturbation theory. Based on the
assumption that the infrared limit of the transverse gluon propagator is
finite,
$- \frac{1}{2} \Pi_{ii}(p_0=0, \vec{p} \to 0) = m_m^2 \sim g^4 T^2$,
one obtains the gauge invariant result~\cite{Re93,Re94}
\begin{equation}
m^2_e(T) = m^2_{e,0}
\left( 1 + \frac{\sqrt{6}}{2 \pi} \, g(T) \frac{m_e}{m_{e,0}} \left[
\log \frac{2 \, m_e}{m_m} - \frac{1}{2} \right] + {\cal O}(g^2) \right) \quad .
\label{rebscreen}
\end{equation}
As the magnetic mass appearing here is expected to be of ${\cal O} (g^2T)$
the next-to-leading order correction is ${\cal O} (g\ln g)$. 

\begin{figure}[htbp]
\begin{center}
  \epsfig{file=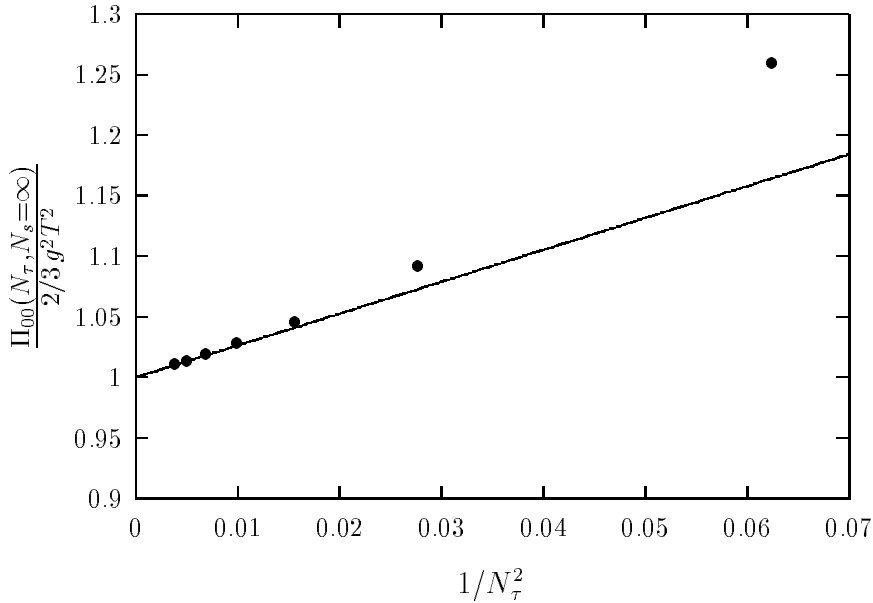, height=80mm}
\end{center}
\caption{The polarization tensor $\Pi_{00}$ at infinite $N_s$ and various
$N_\tau$. The solid line shows the asymptotic result
(\ref{cutofferror}).\label{fig:tensor}}
\end{figure}

It has been pointed out already sometime ago that the leading order
perturbative calculation of the electric screening mass suffers from finite
cut-off effects \cite{ElKaKa88} similarly to what has been known for bulk
thermodynamic observables, e.g.\ the Stefan-Boltzmann law for an ideal gas
\cite{EnKaSa82}. For the Wilson action the leading corrections are
${\cal O}((aT)^2)$, i.e.\ ${\cal O}(N_\tau^{-2})$. This is shown in
Fig.\ \ref{fig:tensor}.
For large $N_\tau$ these deviations are due to the ${\cal O} (a^2)$
discretization errors introduced in the Wilson formulation.  
On a spatially infinite lattice we find for these cut-off errors
\begin{eqnarray}
\frac{\Pi_{00}(N_\tau \to \infty, N_s=\infty)}{2/3 \, g^2 T^2} &=&
1 + N_\tau^{-2} \frac{1}{40 \pi^2} \int_0^\infty \!\!\! dx \: x^5
\frac{\sinh x} {\sinh^4 \frac{x}{2}} \nonumber \\
&=& 1 + {4\over 15} \biggl({\pi \over N_\tau} \biggr)^2 \quad ,
\label{cutofferror}
\end{eqnarray}
which is similar in magnitude to the cut-off dependence of bulk thermodynamic
observables like the energy density \cite{BeKaLa96}. Using an improved action
these leading cut-off errors are eliminated and corrections only start at
${\cal O}(N_\tau^{-4})$. In the case of the energy density or the pressure
these actions lead to a strong reduction of cut-off effects in the high
temperature limit \cite{BeKaLa96}. In the following we will see, however, that
this is not the case for the electric screening mass, i.e.\ the improvement of
the ultra-violet sector does not influence the screening masses much.

\section{Determination of the Temperature Scale}
\label{sec:temperature}
In our simulations we have used both the Wilson action ($S_W$) and a
tree-level Symanzik improved action ($S_I$), 
\begin{eqnarray}
S_W &=& \frac{\beta}{2} \sum_{1 \times 1} \mbox{Tr} \, U_{1 \times 1} \quad ,
\nonumber \\
S_I &=& \frac{\beta}{2} \left(
  \frac{5}{3}  \sum_{1 \times 1} \mbox{Tr} \, U_{1 \times 1}
- \frac{1}{12} \sum_{1 \times 2} \mbox{Tr} \, U_{1 \times 2} \right)
\label{wil_action}
\end{eqnarray}
where the sums run over all elementary $1 \times 1$ plaquettes $U_{1 \times 1}$
and planar $1 \times 2$ Wilson loops $U_{1 \times 2}$~\footnote{In the
following, we will denote the couplings with $\beta_W$ for the Wilson action
and with $\beta_I$ for the Symanzik improved action, respectively.}.
The gauge dependent correlation functions have been calculated in Landau gauge,
$|\partial_\mu A^\mu (x)|^2 = 0$ . Details of our gauge fixing procedure are
given in \cite{HeKaRa95}.

A comparison of simulations performed with these two actions on lattices with
temporal extent $N_\tau=4$ will allow to judge the cut-off dependence of our
results. In the case of the Wilson action we also have performed calculations
on lattices with temporal extent $N_\tau=8$ in order to analyse the cut-off
dependence for this action. 

In order to quantify the influence of the non-zero cut-off at finite
temperature one should, of course, compare calculations at the same physical
temperature, $T\equiv 1/(N_\tau a)$. An accurate determination of the
temperature scale also is needed for analyzing the dependence of the screening
masses on a running coupling, $g(T)$. We thus start our analysis with the
determination of temperature scales for both actions. 

The problem to relate the temperature $T$ to the coupling $\beta$ is equivalent
to the task of finding the dependence of the lattice spacing $a$ on the bare
coupling $g^2$. We follow here the approach outlined in \cite{EnKaRe95}. In
order to take into account the violations of asymptotic scaling in the coupling
regime of interest we use the general ansatz
\begin{eqnarray}
a \, \Lambda_L & = & R(g^2) \cdot \lambda(g^2) \quad \mbox{with}
\label{eq:ansatz} \\
R(g^2) & = & \exp \left[ - \frac{b_1}{2 \, b_0^2} \ln(b_0 g^2) -
\frac{1}{2 \, b_0 g^2} \right] \quad , \\
b_0 = \frac{11 N_c}{48 \, \pi^2} & , &
b_1 = \frac{34}{3} \left( \frac{N_c}{16 \, \pi^2} \right)^2 \quad .
\end{eqnarray}
The function  $\lambda (g^2)$ parameterizes the asymptotic scaling
violations. For this we use an exponential ansatz 
\begin{equation}
\lambda(g^2) = \exp \left[ \frac{1}{2 \, b_0^2} \left(
d_1 g^2 + d_2 g^4 + d_3 g^6 + \ldots \right) \right] \quad .
\label{eq:lambda}
\end{equation}
Using $T = 1 / (N_\tau a) $ we obtain from Eq.\ (\ref{eq:ansatz})
\begin{equation}
\frac{1}{N_\tau R(g_c^2)} = \lambda(g_c^2) \, \frac{T_c}{\Lambda_L} \quad .
\end{equation}
Here $g^2_c$ is the value of the bare coupling at the critical temperature
$T_c$ of the deconfinement phase transition at given $N_\tau$. Using results
for $g_c^2(N_\tau)$ \cite{FiHeKa93,CeCuTrVi94} the function
$\lambda(g_c^2)$ is obtained from a fit where $T_c / \Lambda_L$ is an
additional free parameter.

Based on the Wilson action data for $g^2_c$ summarized in \cite{FiHeKa93},
the best fit in~\cite{EnKaRe95} is given by the parameterization
$d_1 = d_2 = d_{n>3} = 0$. Their fit results are
$d_3 = 5.529(63) \cdot 10^{-4}$ and $(T_c / \Lambda_L)_W = 21.45(14)$.

\begin{figure}[htb]
\begin{center}
  \epsfig{file=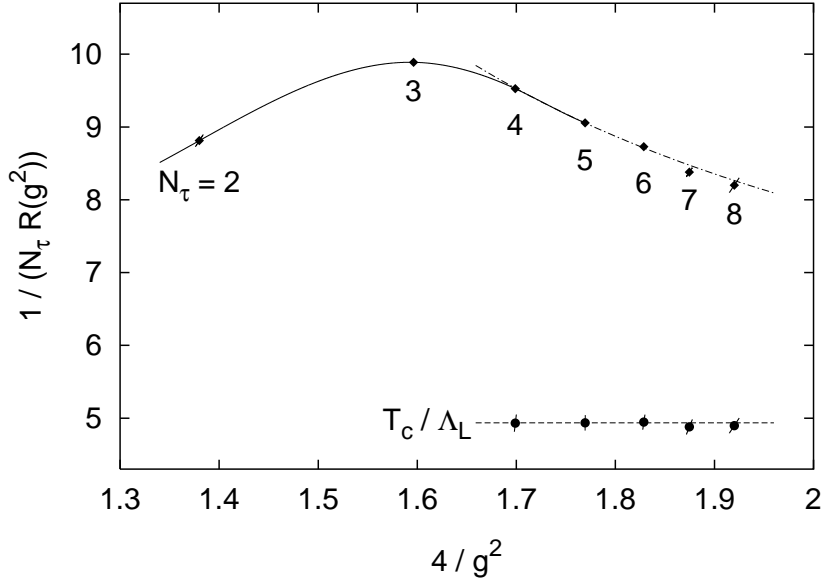, height=80mm}
\end{center}
\caption{The critical temperature $1 / (N_\tau R(g^2))$ vs.\ $4/g^2$ for
$N_\tau = 2,\ldots,8$. The data for $g_c^{-2}(N_\tau)$ are taken from
\cite{CeCuTrVi94}. The solid line is a spline interpolation of the data,
the dashed-dotted line is obtained from a fit for $N_\tau \ge 4$.}
\label{fig:lambda}
\end{figure}

For the Symanzik improved action we have performed a similar fit, using the
critical couplings computed in~\cite{CeCuTrVi94} for $N_\tau \ge 4$. Our best
parameterization is given by $d_1 = d_{n>2} = 0$, and our fit results are
$d_2 = 5.12(18) \cdot 10^{-4}$ and $(T_c / \Lambda_L)_I = 4.94(11)$.
The fit can also be seen in Fig.\ \ref{fig:lambda}.

Using the perturbatively calculated ratios of $\Lambda$-parameter, i.e.
$\Lambda_{L,I}/\Lambda_{L,W} = 4.13089(1)$ \cite{BeWe83WeWo84} and  
$\Lambda_{\overline{\rm MS}}/\Lambda_{L,W} = 19.82314$ \cite{DaGr81}, we 
find for the critical temperature a result which, within 5\%,  coincides with 
the previously determined continuum extrapolation for the Wilson action
\begin{eqnarray}
{T_c \over \Lambda_{\overline{\rm MS}} } & = & \cases{1.08 \pm 0.01 & 
standard Wilson action \cite{EnKaRe95}\cr
1.03 \pm 0.03 & tree level improved (1,2)-action \cr} \quad .
\label{tc}
\end{eqnarray}
In the following, we will use an averaged value of
$T_c / \Lambda_{\overline{\rm MS}} = 1.06$.

We finally need to extract the temperature in units of the critical
temperature at given $N_\tau$. This is given by
\begin{equation}
\frac{T}{T_c} = \frac{R(g_c^2) \cdot \lambda(g_c^2)}
{R(g^2) \cdot \lambda(g^2)} \quad.
\label{eq:t_g}
\end{equation}
Using Eq.\ (\ref{eq:t_g}), the fit results for $\lambda$ and the critical
couplings from~\cite{FiHeKa93,CeCuTrVi94} we can now relate the temperature
$T$ to the coupling $\beta = 4/g^2$. The results for the couplings used in
our analysis are listed in Tab.\ \ref{tab:beta_temp}.
\begin{table}[htb]
\begin{center}
\begin{tabular}{|l|c||l|c||l|c|}
\hline
\multicolumn{2}{|c||}{$N_\tau=4$} & \multicolumn{2}{|c||}{$N_\tau=8$} &
\multicolumn{2}{|c|}{$N_\tau=4$} \\
\hline
\multicolumn{1}{|c|}{$\beta_W$} & $T/T_c$ &
\multicolumn{1}{|c|}{$\beta_W$} & $T/T_c$ & 
\multicolumn{1}{|c|}{$\beta_I$} & $T/T_c$ \\
\hline
2.512 & 2.004 & 2.74 & 2.007 & 1.92  & 1.984 \\
2.643 & 3.002 & 2.88 & 3.031 & 2.063 & 3.031 \\
2.74  & 4.013 & 2.97 & 3.929 & 2.152 & 3.923 \\
2.88  & 6.062 & 3.12 & 6.016 & 2.30  & 5.979 \\
2.955 & 7.527 & 3.20 & 7.530 & 2.382 & 7.528 \\
3.023 & 9.143 & 3.27 & 9.151 & 2.452 & 9.149 \\
3.219 & 15.88 & 3.47 & 15.89 & 2.652 & 15.88 \\
3.743 & 66.78 & 4.00 & 66.71 & 3.183 & 66.68 \\
4.24  & 253.5 & 4.50 & 253.3 & 3.684 & 253.2 \\
4.738 & 953.1 & 5.00 & 953.9 & 4.185 & 954.0 \\
5.238 & 3581  & 5.50 & 3578  & 4.685 & 3572  \\
5.737 & 13383 & 6.00 & 13401 & 5.186 & 13393 \\
\hline
\end{tabular}
\end{center}
\caption{Relations between the couplings and the temperatures.}
\label{tab:beta_temp}
\end{table}
The good agreement found from this analysis for
$T_c / \Lambda_{\overline{\rm MS}}$  calculated with two different actions
suggests that our temperature scale is of similar accuracy.

\section{Screening Masses from Gluon Correlation Functions}
\label{sec:screening_masses}
In our earlier work \cite{HeKaRa95} we have already investigated the
behavior of the electric and magnetic screening masses in Landau gauge.
Whereas in \cite{HeKaRa95} we calculated the gluon propagator only at vanishing
momentum, we now extend the analysis to finite momenta.
Furthermore we are now using temperatures very much higher than in 
\cite{HeKaRa95} in order to possibly
get in closer contact with perturbation theory.
Finally, we are now using in addition to the Wilson action also the tree-level
Symanzik improved action.

In section~\ref{sec:gluonprop} we have given the relations between the energies
in the electric and magnetic sectors and gluonic correlation functions,
Eq.\ (\ref{eq:gluonfunctions}). To extract the screening masses we
use the dispersion relation between energy, screening mass and momentum, which
on the lattice has the form
\begin{equation}
\sinh^2 \frac{a E_i}{2} = \sinh^2 \frac{a m_i}{2} +  \kappa \sum_{j=1}^3 
\sin^2 \frac{a p_j}{2}  \quad , \quad i=e,~m \quad .
\label{eq:disprel}
\end{equation}
In (\ref{eq:disprel}) we have introduced a factor $\kappa$ which parameterizes
deviations from a free particle dispersion relation $(\kappa \equiv 1)$
introduced by a thermal medium.

Using $T = 1/(N_\tau a)$ we can now compute the screening masses in units of
the temperature, $m_i/T$ with $i=e,~m$. We have performed simulations using the
Wilson action on lattices of sizes $32^3 \!\times\! 4$ and
$32^2 \!\times\! 64 \!\times\! 8$ and using the tree-level Symanzik improved
action on a $32^3 \!\times\! 4$ lattice. At each value of the gauge coupling we
performed measurements on at least 1000 configurations, see Tab.\
\ref{tab:measurements}.
\begin{table}[htb]
\begin{center}
\begin{tabular}{|l|c||l|c||l|c|}
\hline
\multicolumn{2}{|c||}{$32^3 \times 4$} &
\multicolumn{2}{|c||}{$32^2 \times 64 \times 8$} &
\multicolumn{2}{|c|}{$32^3 \times 4$} \\
\hline
\multicolumn{1}{|c|}{$\beta_W$} & meas.\ &
\multicolumn{1}{|c|}{$\beta_W$} & meas.\ & 
\multicolumn{1}{|c|}{$\beta_I$} & meas.\ \\
\hline
2.512 &  2000 & 2.74 & 1220 & 1.92  & 2000 \\
2.643 &  2000 & 2.88 & 1000 & 2.063 & 2000 \\
2.74  &  2000 & 2.97 & 1000 & 2.152 & 2000 \\
2.88  &  2000 & 3.12 & 1000 & 2.30  & 2000 \\
2.955 &  2000 & 3.20 & 1000 & 2.382 & 2000 \\
3.023 &  2000 & 3.27 & 1440 & 2.452 & 2000 \\
3.219 &  2000 & 3.47 & 1140 & 2.652 & 2000 \\
3.743 &  2000 & 4.00 & 1000 & 3.183 & 2000 \\
4.24  &  2000 & 4.50 & 1160 & 3.684 & 2000 \\
4.738 &  2000 & 5.00 & 1000 & 4.185 & 2000 \\
5.238 &  2000 & 5.50 & 1000 & 4.685 & 2000 \\
5.737 &  2000 & 6.00 & 1000 & 5.186 & 2000 \\
\hline
\end{tabular}
\end{center}
\caption{Number of measurement.}
\label{tab:measurements}
\end{table}
Two consecutive configurations were separated by at least 10 update iterations,
and each update consists of at least four overrelaxation sweeps, followed by
one heatbath sweep.

From the exponential decay of the gluon correlation functions $G_e, \, G_m$ we
extract the screening masses. A rather technical problem is the procedure to
select a reliable fit range in which $G_e(p_\bot,x_3)$ and $G_m(p_\bot,x_3)$
(see (\ref{eq:gluonfunctions})) can be fitted to extract the energies in the
electric and magnetic sectors. This is described in App.\
\ref{sec:determination}.

The results for the screening masses (from the $\vec{p} = 0$ measurements)
and the energies ($\vec{p} \ne 0$) are listed in
Tabs.\ \ref{tab:m_gluon_648} and~\ref{tab:me_gluon_324} respectively.
\begin{table}[htb]
\begin{center}
\begin{tabular}{|l|l|l||l|l|l|}
\hline
\multicolumn{6}{|c|}{Wilson action, $32^2 \times 64 \times 8$ lattice} \\
\hline\hline
\multicolumn{1}{|c|}{$\beta_W$} &
\multicolumn{1}{c|}{$m_e(T)/T$} & \multicolumn{1}{c||}{$m_m(T)/T$} &
\multicolumn{1}{c|}{$\beta_W$}  &
\multicolumn{1}{c|}{$m_e(T)/T$} & \multicolumn{1}{c|}{$m_m(T)/T$} \\
\hline
2.74 & 2.39(11) & 2.01(29) & 3.47 & 1.62(4) & 0.92(7) \\
2.88 & 1.95(4)  & 1.24(4)  & 4.00 & 1.62(8) & 0.66(3) \\
2.97 & 1.91(7)  & 1.15(4)  & 4.50 & 1.55(5) & 0.61(2) \\
3.12 & 1.92(9)  & 1.23(14) & 5.00 & 1.41(3) & 0.52(3) \\
3.20 & 1.92(10) & 1.09(10) & 5.50 & 1.27(5) & 0.42(2) \\
3.27 & 1.93(6)  & 1.03(5)  & 6.00 & 1.26(5) & 0.37(2) \\
\hline
\end{tabular}
\end{center}
\caption{Electric and magnetic screening masses from $G_e(k_1=0)$ and
$G_m(k_1=0)$.}
\label{tab:m_gluon_648}
\end{table}
\begin{table}[htbp]
\begin{center}
\begin{tabular}{|l|l|l|l|}
\hline
\multicolumn{4}{|c|}{Wilson action, $32^3 \times 4$ lattice} \\
\hline\hline
 & \multicolumn{3}{c|}{$E_e(\vec{p},T)/T$, extracted from} \\
\multicolumn{1}{|c}{$\beta_W$} &
\multicolumn{1}{|c}{$G_e(k_1\!=\!0)$} &
\multicolumn{1}{|c}{$G_e(k_1\!=\!1)$} &
\multicolumn{1}{|c|}{$G_e(k_1\!=\!2)$} \\
\hline
2.512 & 2.14(11) & 2.71(18) & 2.46(5) \\
2.643 & 2.24(9)  & 2.28(8)  & 2.34(4) \\
2.74  & 1.94(5)  & 2.13(5)  & 2.33(7) \\
2.88  & 2.03(7)  & 2.03(4)  & 2.33(5) \\
2.955 & 1.87(4)  & 1.94(4)  & 2.27(5) \\
3.023 & 2.10(14) & 2.12(13) & 2.25(4) \\
3.219 & 1.80(7)  & 1.93(5)  & 2.03(3) \\
3.743 & 1.58(3)  & 1.78(7)  & 1.94(2) \\
4.24  & 1.64(8)  & 1.55(3)  & 2.01(4) \\
4.738 & 1.33(3)  & 1.51(4)  & 1.91(8) \\
5.238 & 1.19(2)  & 1.40(3)  & 1.83(4) \\
5.737 & 1.26(3)  & 1.35(4)  & 1.83(4) \\
\hline
\multicolumn{4}{c}{} \\
\hline
\multicolumn{4}{|c|}{Symanzik action, $32^3 \times 4$ lattice} \\
\hline\hline
 & \multicolumn{3}{c|}{$E_e(\vec{p},T)/T$, extracted from} \\
\multicolumn{1}{|c}{$\beta_I$} &
\multicolumn{1}{|c}{$G_e(k_1\!=\!0)$} &
\multicolumn{1}{|c}{$G_e(k_1\!=\!1)$} &
\multicolumn{1}{|c|}{$G_e(k_1\!=\!2)$} \\
\hline
1.92  & 2.10(6)  & 2.18(5)  & 2.36(4)  \\
2.063 & 1.96(5)  & 2.17(5)  & 2.36(4)  \\
2.152 & 2.08(8)  & 2.04(5)  & 2.23(5)  \\
2.30  & 1.76(3)  & 1.95(4)  & 2.91(26) \\
2.382 & 2.01(11) & 2.00(4)  & 2.54(9)  \\
2.452 & 1.70(5)  & 2.04(9)  & 2.21(5)  \\
2.652 & 1.72(6)  & 1.53(21) & 2.05(3)  \\
3.183 & 1.69(8)  & 1.61(2)  & 2.13(11) \\
3.684 & 1.44(8)  & 1.75(6)  & 1.94(2)  \\
4.185 & 1.50(6)  & 1.30(1)  & 1.85(4)  \\
4.685 & 1.19(6)  & 1.45(7)  & 1.74(2)  \\
5.186 & 1.31(8)  & 1.29(4)  & 1.77(5)  \\
\hline
\end{tabular}
\end{center}
\caption{Energies from the electric sector of gluon correlation functions.}
\label{tab:me_gluon_324}
\end{table}

\subsection{Numerical Results for Zero Momentum}
\label{sec:zero_momentum}
\begin{figure}[htb]
\begin{center}
  \epsfig{file=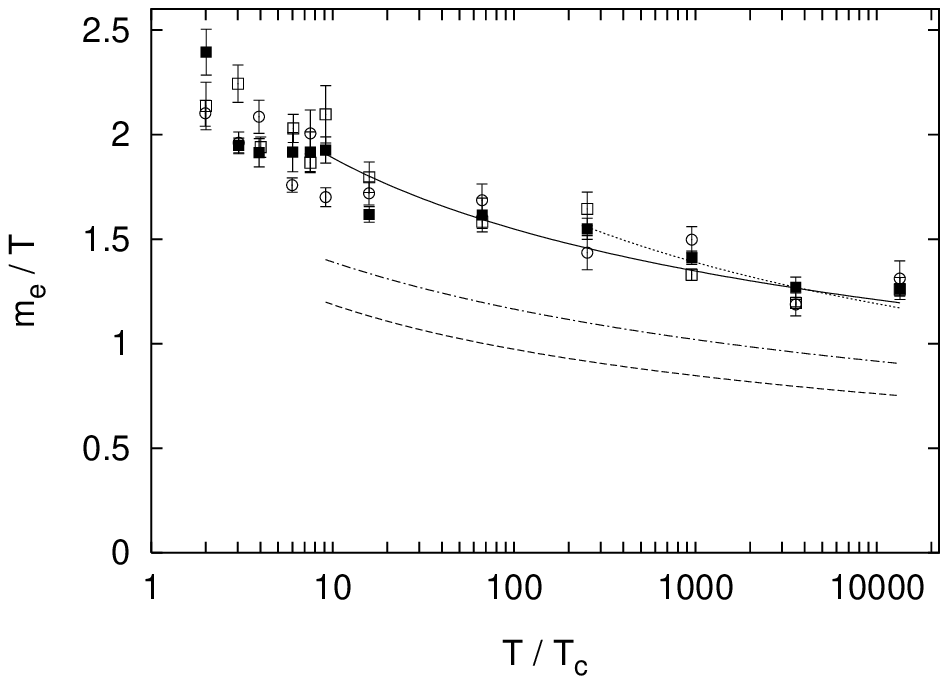, height=80mm}
\end{center}
\vspace{-1.5ex}
\caption{Electric screening masses in units of the temperature vs.\ $T/T_c$
from simulations with Wilson action on $32^2 \!\times\! 64 \!\times\! 8$
(filled squares) and $32^3 \!\times\! 4$ (open squares) lattices and with
Symanzik action on a $32^3 \!\times\! 4$ lattice (open circles). The dashed
line is the tree-level result (\ref{lowestscreen}), the dashed-dotted line is a
self consistent determination of $m_e$, using (\ref{rebscreen}). The other
lines are one parameter fits, using ansatz (\ref{me_fit_1}) (solid line, for
$T \ge 9 \,T_c$) and ansatz (\ref{rebhan_g4_fit}) (dotted line, for
$T \ge 250 \, T_c$) respectively.}
\label{fig:me_gluon}
\end{figure}

Let us first discuss the electric screening mass, extracted from the
measurements at vanishing momentum $\vec{p}=0$. In Fig.\ \ref{fig:me_gluon}
we show $m_e/T$ for both types of actions and the two different lattices
we have used. One can see at once that, within errors,  $m_e/T$ does not
differ significantly for the three sets. Even the tree-level
Symanzik improved action, which cures discretization errors of ${\cal O}(a^2)$
in the action, does not shift the electric screening mass in any direction.
This makes clear that ultra-violet modes do not contribute significantly to the
screening mass. As a consequence, we have analysed all three data sets together.

Fig.\ \ref{fig:me_gluon} shows that $m_e/T$ only depends very weakly on the
temperature for small values of the coupling $\beta$, corresponding to
temperatures less than about $10 \, T_c$. A constant fit in this temperature
range yields $m_e(T)/T =  1.938(15)$. This behavior is qualitatively similar
to what we have observed in \cite{HeKaRa95}. For temperatures 
$1.3 \, T_c < T < 16 \, T_c$ we found in \cite{HeKaRa95}
$m_e(T)/T = 2.484(52)$. The difference between these values arises from
different methods of extracting the screening masses. Whereas in this paper
we performed correlated fits of the gluon correlation functions over variable
fit ranges (see App.\ \ref{sec:determination}), we obtained $m_e$ and $m_m$
in \cite{HeKaRa95} from uncorrelated fits in the fixed range $zT \ge 1$. Our
new method results in screening masses which are up to 20\% smaller. Since
the method accounts for possible correlations in the data the results should
be more reliable.

In contrast to \cite{HeKaRa95} we have measured $m_e$ now also at very
high temperatures (up to $T \sim 13400 \, T_c\,$; see
Tab.\ \ref{tab:beta_temp}). From this analysis it becomes evident that $m_e/T$
runs with $T$. Since this is expected from perturbation theory it is
meaningful to test whether perturbative predictions also work quantitatively.

At lowest order perturbation theory for two color degrees of freedom 
and without taking dynamical quarks into account the electric mass is given
by the well known relation (\ref{lowestscreen}). For the running coupling 
we use the 2-loop formula
\begin{equation}
g^{-2}(T) = \frac{11}{12 \pi^2} \ln \frac{\mu}{\Lambda} +
\frac{17}{44 \pi^2} \ln \left[ 2 \ln \frac{\mu}{\Lambda}
\right]
\end{equation}
with $\mu = 2 \pi T$ being the lowest lying Matsubara frequency. Hence
\begin{equation}
g^{-2}(T) = \frac{11}{12 \pi^2} \left( \ln \frac{T}{T_c} +
\ln \frac{2 \pi T_c}{\Lambda} \right) +
\frac{17}{44 \pi^2} \ln \left[ 2 \left( \ln \frac{T}{T_c} +
\ln \frac{2 \pi T_c}{\Lambda} \right)  \right] \quad .
\label{2_loop_formula}
\end{equation}

We start the discussion of our data with a comparison with (\ref{lowestscreen})
which is shown in Fig.\ \ref{fig:me_gluon} as a dashed line. The numerical
data for $m_e$ are lying about 60\% above the lowest order perturbative result
(\ref{lowestscreen}). However, the functional dependence of the electric mass
on the temperature seems to be well described by $m_e \sim g T$.

To verify the temperature dependence of the electric mass quantitatively we
have performed several fits of $m_e/T$ vs.\ $\ln(T/T_c)$ for temperatures
$T \ge 9 \, T_c$. In our one parameter fits we fix the $\Lambda$-parameter
appearing in the temperature dependent running coupling to
$\Lambda_{\overline{\rm MS}}$ and therefore use the MC-result for
$T_c / \Lambda_{\overline{\rm MS}}\,$, i.e.\
$T_c / \Lambda_{\overline{\rm MS}} = 1.06$. In those cases where we parameterize
the screening masses only by its leading $g^2$ dependence the effect of higher
order corrections can be partially taken into account in a modification of the
$\Lambda$-parameter. We, therefore, also performed two parameter fits with a
free ratio $\Lambda_{\mbox{\scriptsize fit}} / \Lambda_{\overline{\rm MS}}$.

The first fit ansatz we use is
\begin{equation}
\left( \frac{m_e(T)}{T} \right)^2 = A_{\mbox{\scriptsize fit}} \; g^2(T)
\quad .
\label{me_fit_1}
\end{equation}
The results obtained with this ansatz are summarized in 
Tab.\ \ref{tab:g2_gluon_electric}.
\begin{table}[htb]
\begin{center}
\begin{tabular}{|l|c| |l|c|}
\hline
\multicolumn{2}{|c||}{1-parameter fit} & \multicolumn{2}{c|}{2-parameter fit}\\
\hline
$A_{\mbox{\scriptsize fit}}$                                      & 1.69(2)  &
$A_{\mbox{\scriptsize fit}}$                                      & 1.92(9)  \\
$\Lambda_{\mbox{\scriptsize fit}} / \Lambda_{\overline{\rm MS}}$  & 1        &
$\Lambda_{\mbox{\scriptsize fit}} / \Lambda_{\overline{\rm MS}}$  & 0.33(13) \\
$\chi^2 / \mbox{dof}$                    & 4.51     &
$\chi^2 / \mbox{dof}$                    & 4.14     \\
\hline
\end{tabular}
\end{center}
\caption{Fit results of $(m_e(T)/T)^2$, extracted from gluon correlation
functions at zero momentum, using the fit ansatz
(\ref{me_fit_1}).}
\label{tab:g2_gluon_electric}
\end{table}
They again reflect that the lowest order perturbative result
(\ref{lowestscreen}) does not describe the data very well. The fit parameter
$A_{\mbox{\scriptsize fit}}$ is much too big compared to its theoretical value
$2/3$. The solid line shown in Fig.\ \ref{fig:me_gluon} is the result from the
one parameter fit. It shows, as noted above, that at least the variation of
$m_e / T$ with the temperature is well described by ansatz (\ref{me_fit_1}).
However, the temperatures we have used are apparently still too low to get in
contact with lowest order perturbation theory.

To test the next-to-leading order result (\ref{rebscreen}) we also determined
the ratio $m_e / m_m$ and especially the magnetic mass. We were only able to
extract a reliable result for  $m_m$ for the lattice with spatial extension
$N_3 = 64$.  On the smaller lattice the local screening masses  $m_m(x_3,T)$ do
not reach a plateau (see App.\ \ref{sec:determination}). Therefore the fits
of the correlation function $G_m$ were quite poor, i.e.\ had a large $\chi^2$.
As the electric screening masses obtained from different actions and lattice
sizes do not show any significant difference, we expect that also the magnetic
mass does not show a significant ultra-violet cut-off dependence.

\begin{figure}[ptb]
\begin{center}
  \epsfig{file=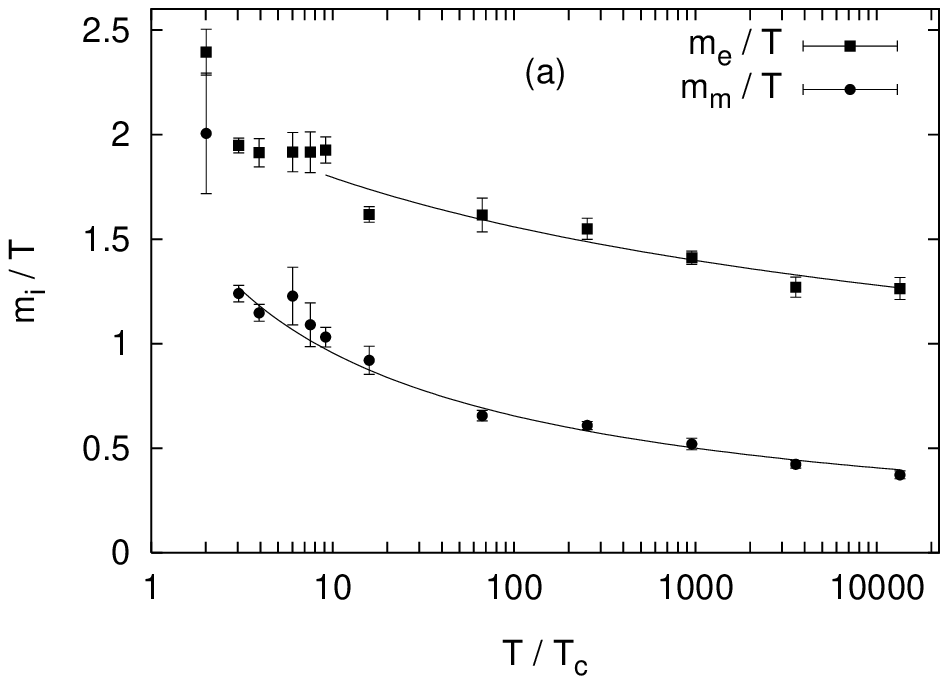, height=90mm}
  \epsfig{file=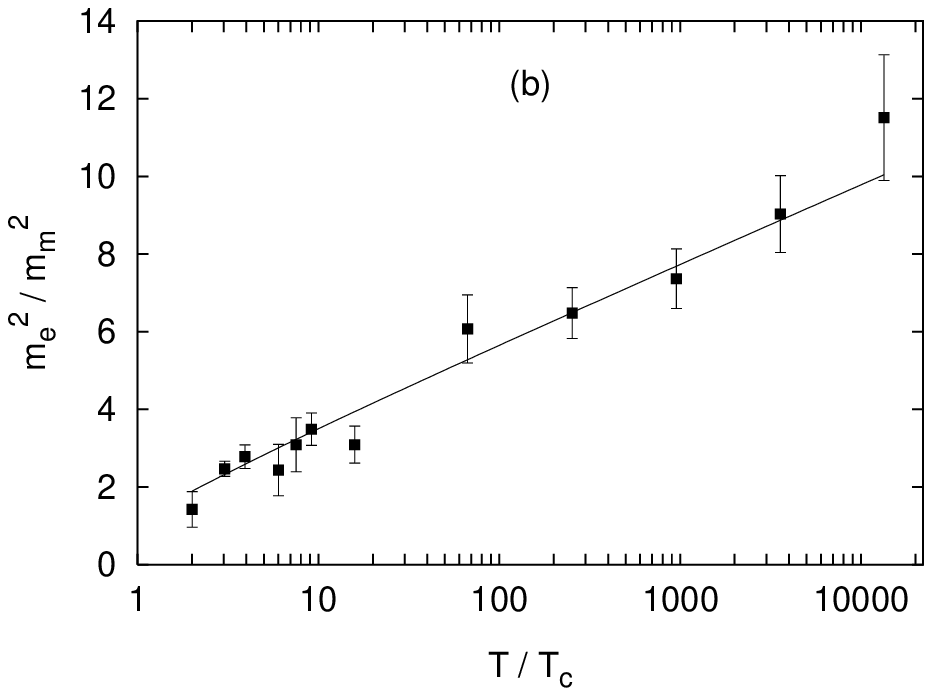, height=90mm}
\end{center}
\caption{Electric and magnetic screening masses in units of the temperature (a)
and squared ratio of the masses (b) vs.\ $T/T_c$. Data are obtained from
simulations on a $32^2 \!\times\! 64 \!\times\! 8$ lattice using the Wilson
action.}
\label{fig:m64x8}
\end{figure}

In Fig.\ \ref{fig:m64x8}a
we show the electric and magnetic screening masses, obtained from the Wilson
action simulation on the $32^2 \times 64 \times 8$ lattice.
Fig.\ \ref{fig:m64x8}b gives the squared ratio $(m_e / m_m)^2$. Our data
strongly suggest a temperature dependence of the form
$(m_e / m_m)^2 \sim g^{-2}(T)$, which is in agreement with the general
expectation $m_m(T) \sim g^2(T) \, T$. We therefore performed a fit according
to 
\begin{equation}
\left( \frac{m_e(T)}{m_m(T)} \right)^2 = C_{\mbox{\scriptsize fit}} \; 
g^{-2}(T)
\quad .
\end{equation}
A two parameter fit in the range $T \ge 2 \, T_c$ yields
$C_{\mbox{\scriptsize fit}} = 9.16(69)$,
$\Lambda_{\mbox{\scriptsize fit}} / \Lambda_{\overline{\rm MS}} = 2.42(64)$
with $\chi^2 / \mbox{dof} = 0.79$. Fixing the $\Lambda$-parameter to
$\Lambda_{\overline{\rm MS}}\,$, a one parameter fit results in
$C_{\mbox{\scriptsize fit}} = 7.46(27)$ and $\chi^2 / \mbox{dof} = 1.35$. In
Fig.\ \ref{fig:m64x8}b we have shown the two parameter fit.

With these results in hand we are now able to check the next-to leading order
result for $m_e$. The dashed-dotted line in Fig.\ \ref{fig:me_gluon} is a
selfconsistent determination of $m_e$, using (\ref{rebscreen}). It lies about
20\% above the lowest order prediction and therefore is closer to our data.
However, it is still too low to describe the data well. Therefore we have
performed additional fits of the electric mass that take into account higher
order corrections. Based on (\ref{rebscreen}) we use the ansatz
\begin{equation}
\left( \frac{m_e(T)}{T} \right)^2 = \frac{2}{3} \, g^2(T) \, 
\left( 1 + \frac{\sqrt{6}}{2 \pi} \, g(T) \,
\left[ \log \frac{2 \, m_e}{m_m} - \frac{1}{2} \right] \right) 
+ B_{\mbox{\scriptsize fit}} \; g^4(T) \quad .
\label{rebhan_g4_fit}
\end{equation}
As the $g^4$ correction term leads to a temperature dependence which is too
strong within the entire $T$-interval, we have restricted the fit to very
high temperatures, $T \ge 250 \, T_c$. A one parameter fit at fixed
$T_c / \Lambda_{\overline{\rm MS}} = 1.06$ gives
$B_{\mbox{\scriptsize fit}} = 0.744(28)$ with $\chi^2 / \mbox{dof} = 4.55$
(dotted line in Fig.\ \ref{fig:me_gluon}).

Let us now return to the discussion of the magnetic mass. As noted above, the
ratio $m_e/m_m$ suggests a magnetic mass of the form $m_m(T) \sim g^2(T) \, T$.
Therefore we fitted $m_m$ with the ansatz
\begin{equation}
\frac{m_m(T)}{T} = D_{\mbox{\scriptsize fit}} \; g^2(T) \quad .
\label{mm_fit}
\end{equation}
The two parameter fit of $m_m$ for $T \ge 3 \, T_c$ results in
$D_{\mbox{\scriptsize fit}} = 0.478(17)$ and
$\Lambda_{\mbox{\scriptsize fit}} / \Lambda_{\overline{\rm MS}} = 0.77(14)$
with $\chi^2  / \mbox{dof}= 1.44$. This is in good agreement with our result
obtained in \cite{HeKaRa95} for $T < 20 \, T_c$. With a fixed
$\Lambda$-parameter, $T_c / \Lambda_{\overline{\rm MS}} = 1.06$,
we obtain $D_{\mbox{\scriptsize fit}} = 0.456(6)$ and
$\chi^2  / \mbox{dof}= 1.53$. In Fig.\ \ref{fig:m64x8}a the two parameter fit
is shown. The small deviation of the fitted curve from the measured data shows
that the magnetic mass indeed is well described by the functional form
$m_m(T) \sim g^2(T) \, T$.

The analysis of $m_m / T$ and $m_e / m_m$ is, of course, consistent with our
most straightforward fit to the electric mass,
$m_e(T) = \sqrt{1.69(2)} \, g(T) \, T$. In contrast to the fit based on
ansatz (\ref{rebhan_g4_fit}) this fit describes the data well in the entire
temperature range above $T_c$. This shows that the electric mass has a strong
non perturbative character in the temperature interval we have investigated.

\subsection{Numerical Results for Non-Zero Momenta}
\label{sec:Finite_momenta}

\begin{figure}[htb]
\begin{center}
  \epsfig{file=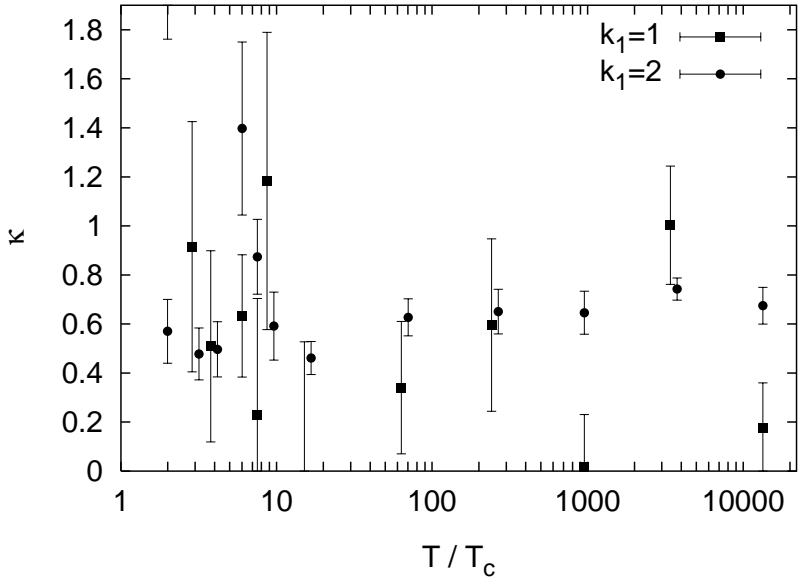, height=80mm}
\end{center}
\caption{$\kappa$ vs.\ $T/T_c$ for $k_1 = 1,2$. Some data points have been
displaced horizontally for better viewing.}
\label{fig:alpha}
\end{figure}

Let us briefly discuss the gluon correlation functions at non-zero
momenta. As the numerical signal gets lost in statistical noise for high
momenta (see (\ref{eq:gluonfunctions}) and (\ref{eq:disprel})) we only could
analyse the cases $k_1 = 1,2$, i.e.\ $p_1 a = 2 \pi / N_1, \, 4 \pi / N_1$.
Furthermore, we only obtained a reliable result in the electric sector.
From Eq.\ (\ref{eq:disprel}) we have
\begin{equation}
\sinh^2 \frac{a E_e(p_1)}{2} = \sinh^2 \frac{a m_e}{2} +  \kappa \,
\sin^2 \frac{a p_1}{2}  \quad .
\label{eq:disprel_el}
\end{equation}
For $m_e$ we use the result from the measurement at zero momentum. In the limit
$T \to \infty$ one expects to find a free particle dispersion relation, i.e.\
$\kappa \to 1$. In Fig.\ \ref{fig:alpha}
we have plotted $\kappa$ vs.\ $T/T_c\,$. Obviously we do not have sufficient
statistics to uncover a temperature dependence of $\kappa$. Therefore we only
quote a value averaged over the temperature interval $T \ge 9 \, T_c$. We find
$\kappa=0.37(10)$ for $k_1 = 1$ and $\kappa=0.65(3)$ for $k_1 = 2$. This
suggest a quite significant modification  of the free particle dispersion
relation at low momenta.

\section{Polyakov Loop Correlation Functions}
\label{sec:polyakov}
It is well known that for temperatures above the critical temperature $T_c$,
the confinement potential between a quark and an anti-quark is replaced by
the color averaged potential \cite{McSv81}, which, in lowest order
perturbation theory, is of the form
\begin{equation}
V_{\mbox{\scriptsize av}} (R,T) \sim \frac{1}{TR^2} \,e^{- 2m_e(T) R}
\quad \mbox{for} \;\, T > T_c\quad .
\label{averaged_potential}
\end{equation}
Here $m_e$ is the electric (or Debye) screening mass. As
$V_{\mbox{\scriptsize av}}$ decreases very fast, the numerical signal gets
lost in statistical noise in the long distance regime. On the other hand,
(\ref{averaged_potential}) is only valid at large distances. This situation
is improved for the color singlet potential, which to leading order
perturbation theory is controlled by 1-gluon exchange and therefore takes on
the form 
\begin{equation}
V_1 (R,T) =  - g^2 \, \frac{N^2_c - 1}{8 \pi N_c} \cdot
\frac{e^{- m_e(T) R}}{R} \quad \mbox{for} \;\, T > T_c \quad .
\label{deconfinement_potential}
\end{equation}
The color singlet potential, however, is gauge dependent and one therefore
again has to fix a gauge for its calculation. As mentioned earlier we have
chosen the Landau gauge.

On the lattice one can extract both potentials by measuring Polyakov loop
correlation functions \cite{McSv81},
\begin{eqnarray}
e^{- V_{\mbox{\scriptsize av}}(R,T)/T} & = &
\frac{ \langle \mbox{Tr} \, L(\vec{R}) \,
\mbox{Tr} \, L^{\dagger}(\vec{0}) \rangle }{ \langle | L | \rangle^2} \quad ,
\label{vav_correlation} \\
e^{- V_1(R,T)/T} & = & 2 \, \frac{ \langle \mbox{Tr} \,( L(\vec{R}) \,
L^{\dagger}(\vec{0})) \rangle }{ \langle | L | \rangle^2} \quad .
\label{v1_correlation}
\end{eqnarray}

\subsection{Improvement of the Singlet Potential}
\label{sec:improvement}
In this section we want to discuss briefly the improvement of rotational
symmetry for the color singlet potential due to the use of an improved
action. We have measured $V_1 / T$ both along an axis,
labeled with $(1,0,0)$, and along three different off-axis
directions, $(1,1,0)$, $(1,1,1)$, and $(2,1,0)$. To make the results from
simulations with unimproved and improved action comparable, one has to
choose couplings that both correspond to the same temperature. As an
example, we use in the following $\beta_W = 3.219$ and $\beta_I = 2.652$.
As listed in Tab.\ \ref{tab:beta_temp}, both couplings correspond to
$T \simeq 15.88 \, T_c$.

\begin{figure}[ptb]
\begin{center}
  \epsfig{file=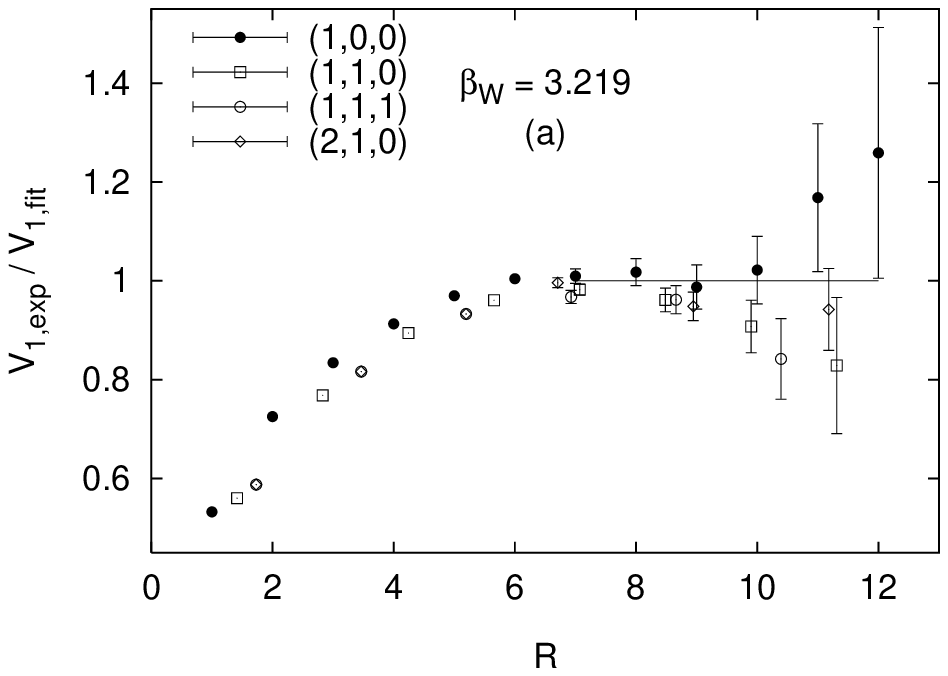, height=90mm}
  \epsfig{file=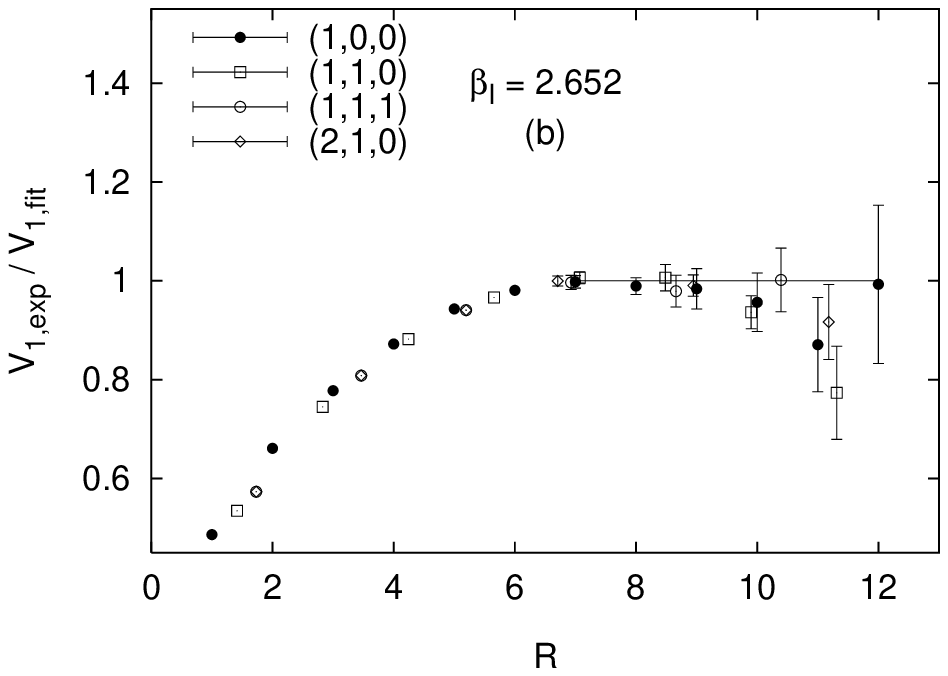, height=90mm}
\end{center}
\caption{Singlet potential $V_1(R)$, normalized by the correlated fit of the
$(1,0,0)$-data in the interval $R \in [7,12]$. The data have been calculated
on a lattice of size $32^3 \!\times\! 4$ using the Wilson action at
$\beta_W=3.219$ (a) and the Symanzik action at $\beta_I=2.652$ (b). Both
couplings correspond to a temperature of $T \simeq 15.88 \, T_c$. The
different symbols refer to the $(x_1,x_2,x_3)-$directions along which the
measurements have been performed.}
\label{fig:v1_wil_imp}
\end{figure}

Motivated by Eq.\ (\ref{deconfinement_potential}) and taking into account
the periodic boundary conditions, we have performed a correlated fit of the
$(1,0,0)$ data in the interval\footnote{See App.\ \ref{sec:determination}.}
$R \in [7,12]$, using the fit function
\begin{equation}
V_{1,\mbox{\scriptsize fit}} (R,T) =  A_{\mbox{\scriptsize fit}} \left( 
\frac{e^{- m_{\mbox{\scriptsize fit}} R}}{R} +
\frac{e^{- m_{\mbox{\scriptsize fit}} (N_3-R)}}{N_3-R} \right)
\label{fit_potential}
\end{equation}
with $N_3=32$. The fit results for both actions are listed in
Tab.\ \ref{tab:v1_compare}.
\begin{table}[htb]
\begin{center}
\begin{tabular}{|c|c|c|}
\hline
fit parameters & $\beta_W = 3.219$ & $\beta_I = 2.652$ \\
\hline
$A_{\mbox{\scriptsize fit}}$ & -1.49(16) & -1.64(13) \\
$m_{\mbox{\scriptsize fit}}$ & 0.435(16) & 0.416(11) \\
goodness                     & 0.380     & 0.452     \\
$\chi^2 / \mbox{dof}$        & 1.049     & 0.918     \\
\hline \hline
\multicolumn{3}{|c|}{normalized $\chi^2$ deviation from the $(1,0,0)$ fit} \\
\hline
$(1,1,0)$                                  & 2.019     & 1.457     \\
$(1,1,1)$                                  & 1.934     & 0.090     \\
$(2,1,0)$                                  & 1.316     & 0.243     \\
\hline
\end{tabular}
\end{center}
\caption{Results from the fits of $V_1/T$ at $\beta_W = 3.219$ and
$\beta_I = 2.652$ $(T \simeq 15.88 \, T_c)$.}
\label{tab:v1_compare}
\end{table}
As one can see from the upper part of the table, the fit itself is better for
the improved data than for the Wilson data, i.e.\ the errors on the fit
parameters are smaller, the goodness is higher and finally the squared error
from the correlated fit ($\chi^2 / \mbox{dof}$) is smaller. The lower part of
Tab.\ \ref{tab:v1_compare} shows the $\chi^2$ deviation of the off-axis
data points from the $(1,0,0)$ fit curves. For this comparison we used data in
the interval $7 \le R \le 12$ and divided by the
number of points taken into consideration. For all measured off-axis
directions these data show that the violation of rotational symmetry is
lowered by going from the Wilson to the tree-level Symanzik improved action.
This behavior becomes also clear from Fig.\ \ref{fig:v1_wil_imp},
which shows the potential $V_1$ (normalized by the fit
function~(\ref{fit_potential}) with the parameters given in
Tab.\ \ref{tab:v1_compare}) vs.\ distance $R \,$.

\subsection{The Electric Screening Mass from the Singlet Potential}
We used on-axis point-to-point as well as plane-plane Polyakov loop correlation
functions to extract the electric screening mass. In the former case, we used
Eqs.\ (\ref{deconfinement_potential}) and (\ref{v1_correlation}) and performed
a correlated fit of the numerical data, using (\ref{fit_potential})
and the fit criterium described in App.\ \ref{sec:determination}. We did
this both for the measurement along the (1,0,0) axis and for the three
different off-axis directions mentioned in Sec.\ \ref{sec:improvement}.

In the second case, the expression for the Polyakov loop $L(\vec{R})$ in
Eq.\ (\ref{v1_correlation}) is replaced by
$L(x_3) \equiv \sum_{x_1,x_2} L(x_1,x_2,x_3)$. Then (\ref{v1_correlation}) and
(\ref{deconfinement_potential}) transform into
\begin{equation}
e^{- V_{1,\mbox{\scriptsize sum}}(x_3,T)/T} = 2 \; \frac{ \langle
\mbox{Tr} \,( L(x_3) \, L^{\dagger}(0)) \rangle }{ \langle | L | \rangle^2}
\label{v1sum_correlation}
\end{equation}
and
\begin{equation}
V_{1,\mbox{\scriptsize sum}}(x_3,T) \sim e^{- m_e(T) x_3} 
\quad \mbox{for} \;\, T > T_c \quad .
\label{sum_deconfinement_potential}
\end{equation}
Whereas we have calculated $V_1$ on lattices of size $32^3 \!\times\! 4$ and
$32^2 \!\times\! 64 \!\times\! 8$ and for both actions, we have calculated
$V_{1,\mbox{\scriptsize sum}}$ only on the larger lattice, using the Wilson
action. The results for the electric screening mass  are listed in
Tab.\ \ref{tab:me_polyakov_324} and \ref{tab:me_polyakov_648}.
\begin{table}[htbp]
\begin{center}
\begin{tabular}{|l|l|l|l|l|l|}
\hline
\multicolumn{6}{|c|}{Wilson action, $32^3 \times 4$ lattice} \\
\hline\hline
 & \multicolumn{5}{|c|}{$m_e(T)/T$, extracted from} \\
\multicolumn{1}{|c|}{$\beta_W$} &
\multicolumn{1}{|c|}{$V_{1,\mbox{\scriptsize sum}}$} &
\multicolumn{1}{|c|}{$V_{1,(1,0,0)}$} & \multicolumn{1}{|c|}{$V_{1,(1,1,0)}$} &
\multicolumn{1}{|c|}{$V_{1,(1,1,1)}$} & \multicolumn{1}{|c|}{$V_{1,(2,1,0)}$} \\
\hline
2.512 & 2.03(2)  & 2.28(9)  & 2.22(4) & 2.13(4) & 2.38(11) \\
2.643 & 2.30(9)  & 2.15(6)  & 2.12(5) & 2.07(4) & 2.25(9)  \\
2.74  & 2.13(9)  & 2.09(5)  & 2.04(5) & 2.04(4) & 2.14(7)  \\
2.88  & 2.04(6)  & 1.95(6)  & 1.91(3) & 1.93(2) & 1.96(7)  \\
2.955 & 1.94(7)  & 2.11(8)  & 1.97(5) & 1.94(6) & 2.06(5)  \\
3.023 & 2.16(7)  & 1.84(5)  & 2.01(4) & 1.96(9) & 1.98(6)  \\
3.219 & 1.88(9)  & 1.74(6)  & 1.80(4) & 1.75(5) & 1.83(4)  \\
3.743 & 1.85(15) & 1.50(2)  & 1.50(1) & 1.43(2) & 1.61(3)  \\
4.24  & 1.63(7)  & 1.55(9)  & 1.53(7) & 1.45(3) & 1.44(2)  \\
4.738 & 1.33(4)  & 1.75(10) & 1.32(5) & 1.32(3) & 1.29(2)  \\
5.238 & 1.30(7)  & 1.29(4)  & 1.25(4) & 1.20(2) & 1.19(2)  \\
5.737 & 1.34(5)  & 1.29(3)  & 1.24(4) & 1.17(1) & 1.15(2)  \\
\hline
\multicolumn{6}{c}{} \\
\hline
\multicolumn{6}{|c|}{Symanzik action, $32^3 \times 4$ lattice} \\
\hline\hline
 & \multicolumn{5}{|c|}{$m_e(T)/T$, extracted from} \\
\multicolumn{1}{|c|}{$\beta_I$} &
\multicolumn{1}{|c|}{$V_{1,\mbox{\scriptsize sum}}$} &
\multicolumn{1}{|c|}{$V_{1,(1,0,0)}$} & \multicolumn{1}{|c|}{$V_{1,(1,1,0)}$} &
\multicolumn{1}{|c|}{$V_{1,(1,1,1)}$} & \multicolumn{1}{|c|}{$V_{1,(2,1,0)}$} \\
\hline
1.92  & 2.26(9)  & 2.04(4)  & 2.16(5)  & 2.12(5) & 2.06(2)  \\
2.063 & 1.97(3)  & 1.98(3)  & 2.01(3)  & 2.03(4) & 2.12(7)  \\
2.152 & 2.07(5)  & 2.16(8)  & 2.02(7)  & 2.04(4) & 2.01(6)  \\
2.30  & 1.93(5)  & 1.83(2)  & 1.79(3)  & 1.70(3) & 1.84(4)  \\
2.382 & 1.82(4)  & 1.83(3)  & 1.72(3)  & 1.70(2) & 1.90(5)  \\
2.452 & 1.78(6)  & 2.14(12) & 1.98(14) & 1.76(7) & 1.79(5)  \\
2.652 & 1.69(5)  & 1.66(4)  & 1.57(2)  & 1.54(2) & 1.67(3)  \\
3.183 & 1.65(6)  & 1.58(6)  & 1.62(5)  & 1.51(3) & 1.48(2)  \\
3.684 & 1.38(4)  & 1.71(12) & 1.42(3)  & 1.36(3) & 1.33(2)  \\
4.185 & 1.62(7)  & 1.66(15) & 1.34(3)  & 1.19(2) & 1.21(66) \\
4.685 & 1.46(12) & 1.22(3)  & 1.19(3)  & 1.10(2) & 1.09(1)  \\
5.186 & 1.16(2)  & 1.22(5)  & 1.14(3)  & 1.05(2) & 1.03(1)  \\
\hline
\end{tabular}
\end{center}
\caption{Electric screening masses from Polyakov loop correlation functions.}
\label{tab:me_polyakov_324}
\end{table}
\begin{table}[htbp]
\begin{center}
\begin{tabular}{|l|l||l|c||l|l|}
\hline
\multicolumn{6}{|c|}{Wilson action, $32^2 \times 64 \times 8$ lattice} \\
\hline\hline
\multicolumn{1}{|c|}{$\beta_W$} & \multicolumn{1}{c||}{$m_e(T)/T$} &
\multicolumn{1}{c|}{$\beta_W$}  & \multicolumn{1}{c||}{$m_e(T)/T$} &
\multicolumn{1}{c|}{$\beta_W$}  & \multicolumn{1}{c|}{$m_e(T)/T$} \\
\hline
2.74 & 2.46(16) & 3.20 & 1.76(4)  & 4.50 & 1.55(10) \\
2.88 & 1.89(4)  & 3.27 & 1.93(5)  & 5.00 & 1.36(3)  \\
2.97 & 1.80(31) & 3.47 & 1.65(3)  & 5.50 & 1.18(8)  \\
3.12 & 1.92(8)  & 4.00 & 1.61(4)  & 6.00 & 1.16(31) \\
\hline
\end{tabular}
\end{center}
\caption{Electric screening masses from $V_{1,\mbox{\scriptsize sum}}$.}
\label{tab:me_polyakov_648}
\end{table}

\begin{figure}[tb]
\begin{center}
  \epsfig{file=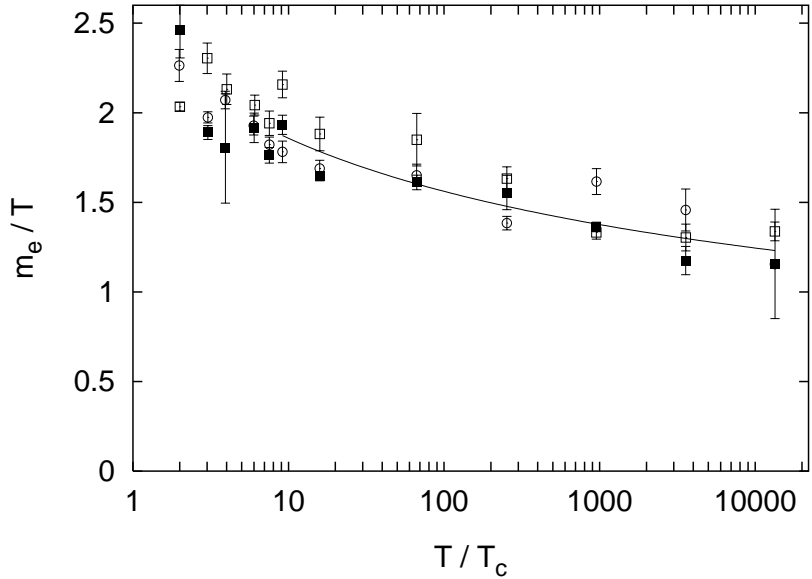, height=80mm}
\end{center}
\caption{Electric screening masses, obtained from
$V_{1,\mbox{\scriptsize sum}}$. The data points refer to the same lattice sizes
and actions as in Fig.\ \ref{fig:me_gluon}. The solid line is a one parameter
fit for $T \ge 9 \,T_c$, using ansatz (\ref{me_fit_1}).}
\label{fig:me_polyakov}
\end{figure}

Similar to the electric mass extracted from gluon correlation functions, the
results we have now obtained with different actions and on lattices of varying
size again do not differ significantly. Therefore we have also here analysed
all three datasets together. The screening masses, extracted from
$V_{1,\mbox{\scriptsize sum}}$, are shown in Fig.\ \ref{fig:me_polyakov}.

As expected from Sec.\ \ref{sec:screening_masses}, $m_e/T$, extracted now
from Polyakov loop correlation functions, also depends only weakly on the
temperature for temperatures less than about $9 \, T_c$. For example, fitting
$V_{1,\mbox{\scriptsize sum}}$ in this temperature range with a constant, we
obtain $m_e(T)/T = 2.010(13)$. 

According to Sec.\ \ref{sec:screening_masses} we performed, for temperatures
$T \ge 9 \, T_c$, one parameter fits
(i.e.\ $\Lambda_{\mbox{\scriptsize fit}} / \Lambda_{\overline{\rm MS}} = 1$),
using the ansatz (\ref{me_fit_1}). The results from the lattice of size
$32^3 \!\times\! 4$ are summarized in Tab.\ \ref{tab:g2_polyakov}.
\begin{table}[htb]
\begin{center}
\begin{tabular}{|l| |c|c|c|c|c|}
\hline
\multicolumn{6}{|c|}{Wilson action, $32^3 \times 4$ lattice} \\
\hline\hline
 & \multicolumn{5}{|c|}{Fits of $(m_e(T)/T)^2$, extracted from} \\
 & $V_{1,\mbox{\scriptsize sum}}$ & $V_{1,(1,0,0)}$ & $V_{1,(1,1,0)}$ &
   $V_{1,(1,1,1)}$ & $V_{1,(2,1,0)}$ \\
\hline\hline
$A_{\mbox{\scriptsize fit}}$ &
 1.97(6) & 1.70(3) & 1.60(2) & 1.53(2) & 1.61(2) \\
$\chi^2 / \mbox{dof}$        &
 2.49    & 10.6    & 6.64    & 7.29    & 4.18    \\
\hline
\multicolumn{6}{c}{} \\
\hline
\multicolumn{6}{|c|}{Symanzik action, $32^3 \times 4$ lattice} \\
\hline\hline
 & \multicolumn{5}{|c|}{Fits of $(m_e(T)/T)^2$, extracted from} \\
 & $V_{1,\mbox{\scriptsize sum}}$ & $V_{1,(1,0,0)}$ & $V_{1,(1,1,0)}$ &
   $V_{1,(1,1,1)}$ & $V_{1,(2,1,0)}$ \\
\hline\hline
$A_{\mbox{\scriptsize fit}}$ &
 1.67(4) & 1.46(2) & 1.46(2) & 1.35(2) & 1.36(2) \\
$\chi^2 / \mbox{dof}$        &
 7.10    & 15.0    & 10.8    & 4.51    & 5.72    \\
\hline
\end{tabular}
\end{center}
\caption{Fit results of $(m_e(T)/T)^2$, extracted from Polyakov loop
correlation functions, using the fit ansatz (\ref{me_fit_1}).}
\label{tab:g2_polyakov}
\end{table}
On the $32^2 \!\times\! 64 \!\times\! 8$ we obtain from 
$V_{1,\mbox{\scriptsize sum}}$ a fit value $A_{\mbox{\scriptsize fit}}=1.72(4)$
with $\chi^2 / \mbox{dof} = 4.60$.

In general we find that the results extracted from
$V_{1,\mbox{\scriptsize sum}}$ are in good agreement with the zero momentum
results from the gluon correlation functions. To make this clear also
quantitatively we have analyzed all three datasets for
$V_{1,\mbox{\scriptsize sum}}$ together, as in the case of the gluon
correlation functions. The one parameter fit for $T \ge 9 T_c$ yields
$A_{\mbox{\scriptsize fit}} = 1.71(2)$ with $\chi^2 / \mbox{dof} = 5.80$.
This can be compared with the result from Tab.\ \ref{tab:g2_gluon_electric},
$A_{\mbox{\scriptsize fit}} = 1.69(2)$ with $\chi^2 / \mbox{dof} = 4.51$.
We therefore conclude that the electric screening mass is well described by
$m_e(T) = \sqrt{1.70(2)} \, g(T) \, T$ in the temperature range
$T \le 14000 \, T_c$.

\section{Summary and Conclusions}
\label{sec:summary}
We have studied Polyakov loop and gluon correlation functions in the high
temperature deconfined phase of SU(2) lattice gauge theory in a wide range
of temperatures, using both the standard Wilson action and a tree-level
Symanzik improved action. We have calculated chromo-electric and -magnetic
screening masses and have determined their dependence on the temperature. 

The temperature dependence found for the magnetic mass is in accordance with
the expected $g^2(T)$-de\-pend\-ence. We find
$m_m(T) = 0.456(6) \, g^2(T) \, T$. The behavior of the electric mass shows,
however, that this does not at all mean that the screening masses can be
described according to perturbative expections. Although the temperature
dependence of $m_e$ is consistent with a logarithmic
dependence, $m_e \sim gT$, our data do not agree with lowest order perturbation
theory. Only little improvement is achieved by using next-to-leading order
results from resummed PT. From an analysis of the gluon propagator as well as
the color singlet potential we find $m_e(T) = \sqrt{1.70(2)} \, g(T) \, T$.
This result shows that the screening mechanism is highly non-perturbative even
for temperatures as large as $14000 \, T_c$. This observation is in accordance
with studies of screening in dimensionally reduced 3d-QCD
\cite{KaLaRuSh97,KaLaPeRaRuSh97}.

Our simulation of the gluon correlation functions at finite momenta still
suffer from insufficient statistics. We find a modification of the energy
momentum dispersion relation of a free particle, but we are not yet able to
quantify its temperature dependence.

The improvement of the action does not show, within statistical errors, any 
significant modification of the behavior of the screening masses, although we
can show that the violation of the rotational symmetry of the singlet
potential, which also was used to extract $m_e$, is weakened.

\medskip
\noindent
{\bf Acknowledgements:}
The work of FK has been supported through the Deutsche
Forschungsgemeinschaft under grant Pe 340/3-3 and the work of UMH and
JR, in part, by the US DOE
under grants DE-FG05-85ER25000 and DE-FG05-96ER40979. JR acknowledges
a graduate scholarship of North-Rhine-Westfalia and support by the DAAD.

\vspace{2em}
\begin{appendix}
\noindent
{\bf \LARGE Appendix}
\section{Determination of Screening Masses from Correlation Functions}
\label{sec:determination}
To shorten the discussion, we assume in this section the case of vanishing
momentum. Taking periodic boundary conditions into account,
it follows from Eqs.\ (\ref{eq:gluonfunctions}) and
(\ref{eq:disprel}) that the electric and magnetic screening masses
are related to the long-distance behavior of the gauge field correlation
functions $G_e(x_3) \equiv G_e(p_\bot=\!0,x_3)$ and
$G_m(x_3) \equiv G_m(p_\bot=\!0,x_3)$ via
\begin{equation}
G_i(x_3) \sim \cosh \left\{ m_i \left(x_3 - \frac{N_3}{2} \right) \right\}
\quad \mbox{for} \; x_3 \gg 1 \; \mbox{and} \; i=e,m \, .
\end{equation}
As there is generally no unique rule how to select the interval in which
$G_e(x_3)$ and $G_m(x_3)$ should be fitted, we have constructed a criterium how
to find a suitable fit interval for a given set of numerical data.

A fit is in general characterized by several properties, namely $\chi^2$,
goodness ($Q$), degrees of freedom ($\varrho$) and relative errors of the fit
parameters. As $\chi^2$ enters directly into the calculation of $Q$,
it is sufficient to consider only the last three quantities. It is desirable
to find a fit with large $Q$, large $\varrho$ and small relative error on the
fit parameter we are interested in, i.e.\ on $m_i/T$. As we want to weigh
these three quantities, we are looking for a fit interval with
\begin{equation}
Q^{\alpha} \cdot \varrho^{\beta} \cdot
\left( \frac{\Delta m_i}{m_i} \right)^{-\gamma} \to \max \quad .
\label{fit_condition}
\end{equation}
We have chosen the coefficients to be $\alpha = 9$, $\beta = 1$ and
$\gamma = 3$. As we put the largest weight on $Q$, it sometimes happens that
only a very small fit interval is selected by this condition. To avoid this
problem, we add the extra condition $\varrho \ge 3$. For a two parameter fit
this is equivalent to demand that the fit interval should contain at least 5
points.

Finally we require that the fit only considers points $G_e(x_3)$ or
$G_m(x_3)$ with errors less than 50 \% of their value. This is to reject
points that are dominated by statistical noise.

\begin{figure}[htb]
\begin{center}
  \epsfig{file=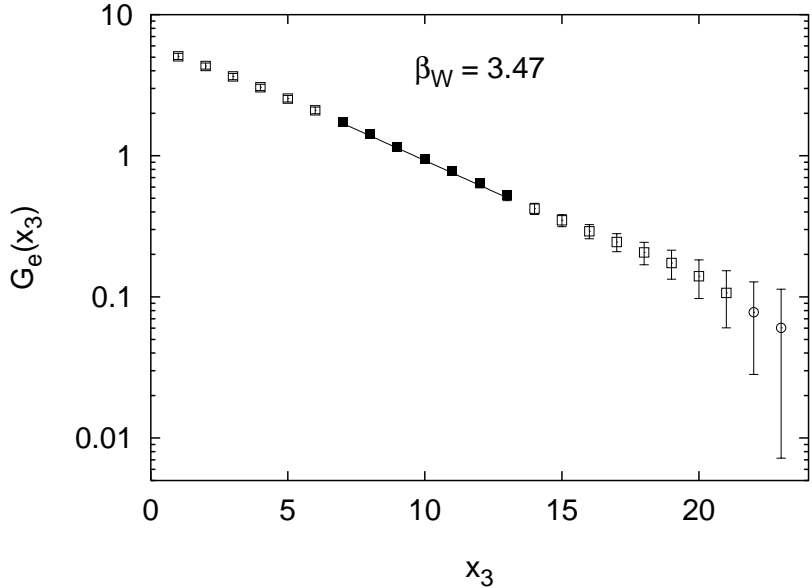, height=80mm}
\end{center}
\caption{The electric correlation function $G_e(x_3)$ as a function of $x_3$
calculated on a lattice of size $32^2 \!\times\! 64 \!\times\! 8$ using the
Wilson action at $\beta_W=3.47$. The squares show points with an error less
than 50\% of the value, whereas the circles describe points with a bigger
error. The filled points represent the fit interval, found by our fit
criterium. The solid line is the correlated fit.}
\label{fig:pe_local}
\end{figure}

To see how our fit criterium works we show in Fig.\ \ref{fig:pe_local}
$G_e(x_3)$ vs.\ $x_3$ for one arbitrary coupling, $\beta_W = 3.47$. The solid
line is the correlated fit, found automatically
by the fit criterium. To demonstrate the quality of the fit criterium, we have
also studied local screening masses $m_i(x_3)$. They are defined by the
relation ($a=1$)
\begin{equation}
\frac{G_i(x_3)}{G_i(x_3+1)} =
\frac{{\rm cosh} \left( m_i(x_3) \left( x_3 - \frac{N_3}{2}
\right) \right)}
{{\rm cosh} \left( m_i(x_3) \left( x_3 + 1 - \frac{N_3}{2}
\right) \right)} \quad , \quad i=e,~m \quad .
\label{mratios}
\end{equation}
If $x_3$ becomes large enough, $m_i(x_3)$ must reach a plateau. On the other
hand, if $x_3$ becomes too large, the local masses have big statistical errors
and do not carry valuable information.

\begin{figure}[tbh]
\begin{center}
  \epsfig{file=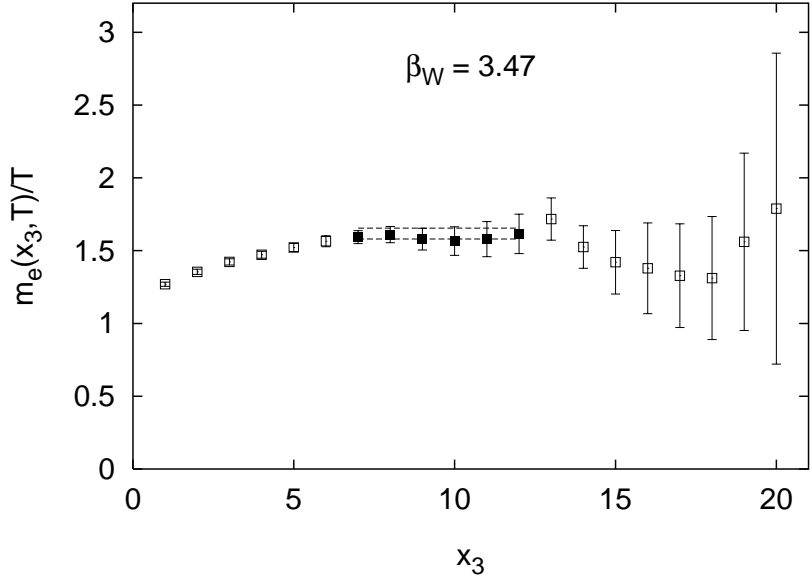, height=80mm}
\end{center}
\caption{Local electric screening masses, extracted from the electric
correlation function $G_e(x_3)$ shown in Fig.\ \ref{fig:pe_local}.
The horizontal lines are the lower and upper bounds for $m_e(T)/T$, given
by the correlated fit shown in Fig.\ \ref{fig:pe_local}.}
\label{fig:m_local}
\end{figure}

In Fig.\ \ref{fig:m_local}
we show the local electric screening masses in units of the temperature,
$m_e(x_3,T) / T$, extracted from $G_e(x_3)$ shown in
Fig.\ \ref{fig:pe_local}. The horizontal lines are the lower and upper bounds
for $m_e(T)/T$, given by the correlated fit shown in Fig.\ \ref{fig:pe_local}.
Obviously, the fit criterium works fine and finds a reliable fit interval.

\end{appendix}

\end{document}